\documentclass[aps,pra,twocolumn,showpacs,groupedaddress,superscriptaddress,longbibliography]{revtex4-1}
\usepackage[english]{babel}
\usepackage{float}
\usepackage{graphicx}
\usepackage[caption=false]{subfig}
\usepackage{amsmath}
\usepackage{amsfonts}
\usepackage{amssymb}
\usepackage{mathbrack}
\usepackage{mathtools}
\usepackage{siunitx}
\usepackage{mathtools}

\begin{document}

\title{Degenerate subspace localization and local symmetries}

\author{Peter Schmelcher}
\email{Peter.Schmelcher@physnet.uni-hamburg.de}
\affiliation{Zentrum f\"ur Optische Quantentechnologien, Fachbereich Physik, Universit\"at Hamburg, Luruper Chaussee 149, 22761 Hamburg, Germany}
\affiliation{The Hamburg Centre for Ultrafast Imaging, Universit\"at Hamburg, Luruper Chaussee 149, 22761 Hamburg, Germany}

\date{\today}

\begin{abstract}
Domain specific localization of eigenstates has been a persistent observation for systems with local
symmetries. The underlying mechanism for this localization behaviour has however remained elusive.
We provide here an analysis of locally reflection symmetric tight-binding Hamiltonian which attempts at identifying
the key features that lead to the localized eigenstates. A weak coupling expansion of closed-form
expressions for the eigenvectors demonstrates that the degeneracy of on-site energies occuring at the center of the
locally symmetric domains represents the nucleus for eigenstates spreading across the domain. 
Since the symmetry-related subdomains constituting a locally symmetric domain are isospectral
we encounter pairwise degenerate eigenvalues that split linearly with an increasing coupling strength of the subdomains.
The coupling to the (non-symmetric) environment
in an extended setup then leads to the survival of a certain system specific fraction of linearly
splitting eigenvalues. The latter go hand in hand with the eigenstate localization on the
locally symmetric domain. We provide a brief outlook addressing possible generalizations
of local symmetry transformations while maintaining isospectrality.

\end{abstract}

\maketitle

\section{Introduction} \label{sec:introduction}

\noindent
The formation of ordered phases of matter is of major importance for the characterization
and understanding of complex quantum systems. This includes highly ordered microscopic structures
such as crystal lattices \cite{Bruus04} as well as synthetically prepared quantum matter based e.g. on ultracold
neutral atoms in optical lattices \cite{Bloch08} or highly excited Rydberg atoms in arrays of optical tweezers 
\cite{Browaeys20}. Symmetries play a crucial role in what state a system adopts ranging from strongly
ordered to completely disordered. In case of geometrical symmetries each of the well-known symmetries,
such as rotation, translation, reflection or inversion leave their characteristic fingerprints in the
corresponding spectral and eigenstate properties of the underlying quantum system. For global symmetries,
i.e. if a symmetry holds for the complete system under investigation, the corresponding group theoretical
representation \cite{Hamermesh89} allows us to make powerful predictions beyond the case of
system-specific computational studies. If a symmetry is broken globally but retained locally the
powerful toolbox of global symmetry analysis does not apply. The corresponding symmetry operations do not
commute with the Hamiltonian since the part of the system where the local symmetry holds is embedded into and 
coupled to the complementary part of the total system.

\noindent
Naturally occuring or synthetically prepared setups with local symmetries represent a bridge between
global order and disorder. Aperiodic long-range order occuring in quasicrystals fall into the 
mentioned gap and exhibit a plethora of spatially-varying local symmetries 
\cite{Macia09,Macia21,Shechtman84,Suck02,Janssen86,Berger93,Vieira05,Tanese14,Jagannathan21,Morfonios14}.
Their arrangement is responsible for 
novel physical properties such as the fractal nature of the energy spectra and the critical localization
of the eigenstates. Energy eigenvalues can cluster in so-called quasibands 
\cite{Prunele01,Prunele02,Bandres16,Vignolo16,Macia17} that are susceptible to
deviations from the perfect aperiodic long-range order. They can develop edge states
in case of a finite quasiperiodic sequence \cite{Roentgen19}
which are, in general, remnants of their topological order \cite{Prunele02,Bandres16}.

\begin{figure*}[t]
\centering
\includegraphics[width=\linewidth]{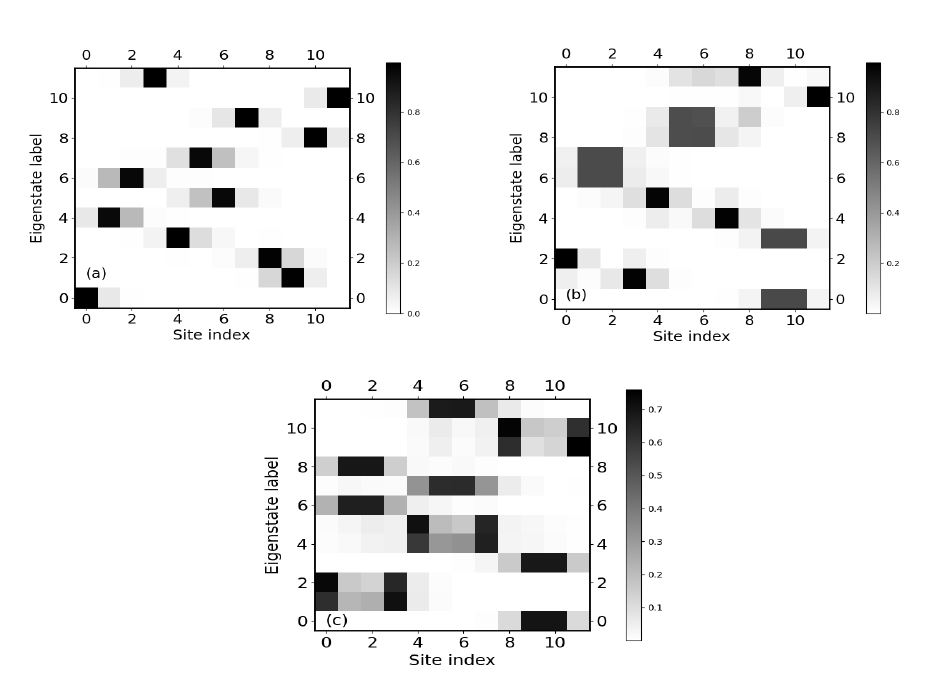}
\caption{Eigenstate maps showing the absolute values of the eigenvector components with varying site index (horizontal axis)
for an increasing degree of excitation (vertical axis), i.e. an increasing energy,
on grey scale for different 12-dimensional TB Hamiltonians.
(a) Absence of local symmetries. Diagonal values of the TB-Hamiltonian are $0.8,2.4,2.9,5.0,1.9,3.0,2.5,4.0,1.8,0.9,3.1,4.9$
and the off-diagonal coupling value is $\epsilon=0.15$. (b) Three consecutive 4-dimensional domains of local reflection
symmetry constituting a 12-dimensional TB Hamiltonian for $\epsilon=0.15$. The diagonal values are
$0.8,2.4,2.4,0.8,1.9,3.0,3.0,1.9,3.2,0.9,0.9,3.2$, i.e. the LS-domains reside on the sites
$0-3$, $4-7$ and $8-11$, respectively. (c) Same diagonal values as in (b) but for $\epsilon = 0.45$ for
all intradomain couplings and $\epsilon=0.1$ for all interdomain couplings.}
\label{Fig1}
\end{figure*}

\noindent
The consequences of the presence of local symmetries for the theoretical description and simulation
of corresponding setups have been pursued intensely in the past decade. In spite of the broken global
symmetry the presence of the local symmetry leads to the existence of invariant non-local currents that
generalize the parity and Bloch theorem \cite{Kalozoumis14a,Schmelcher17,Morfonios17} for the case of
reflection and translation symmetries, respectively. These invariants and their impact on the
structure and dynamics have been verified experimentally for both acoustic \cite{Kalozoumis15} 
and optical \cite{Schmitt20} wave propagation. One of the resulting applications is the classification
of scattering resonances based on sum rules for the invariants \cite{Kalozoumis13}. Systematically
introducing more and more local symmetries into an originally disordered chain has been shown
to enhance the corresponding transfer efficiency across the chain \cite{Morfonios20}. An important
and always recurring theme and feature of corresponding devices with (many) local symmetries
is their eigenstate localization properties on the domains with local symmetries, i.e.
LS-domains, in the total system. 
Specifically a real space local resonator approach \cite{Roentgen19} has been developed for binary tight-binding (TB) chains with
aperiodic long-range order applying to the situation of sufficiently strong contrast i.e. weak to 
intermediate coupling. This approach can be used to predict the occurrence of gap-edge states and the design of
their spectral occurence. Overlapping local symmetries and consequently the 'flexible' localization 
of eigenstates does indeed lead to the bridging of transport across the device \cite{Morfonios20}.

\noindent
To exemplify the above statements on the interrelation between the presence of local symmetries and the
localization properties of the resulting eigenstates we show in Fig.\ref{Fig1} the eigenstate maps
of the absolute values of the eigenvector components for different 12-dimensional TB
Hamiltonians. Fig. \ref{Fig1}(a) shows the case of no local symmetry being present in the Hamiltonian.
Obviously for the shown small values of the coupling $\epsilon$ each of the states is strongly
localized on its parental zero-coupling site with some smaller amplitude on the left and right-localized
neighboring sites. Fig. \ref{Fig1}(b) represents the eigenstate map for a setup consisting of three coupled
4-dimensional LS-domains which reside on the sites $0-3$ with on-site values $0.8,2.4,2.4,0.8$, sites $4-7$ with
values $1.9,3.0,3.0,1.9$ and sites $8-11$ with values $3.2,0.9,0.9,3.2$,
respectively. These LS-domains possess a (local) reflection symmetry and each consists of two subdomains
which are transformed onto each other by the reflection mapping. As a consequence there exist
two neighboring central sites for each LS-domain which possess the same on-site energies.
The important observation concerning Fig.\ref{Fig1}(b) is now that six of the eigenstates show a profile which is distinctly
localized on the LS-domains whereas the profile of the others is reminescent of what
we observe in the non-symmetric case of Fig.\ref{Fig1}(a). The local symmetry localized states are not
necessarily energetically neighboring states (see the 7-th and 8-th versus the 1-st and 4-th eigenstates).

\noindent
Finally, Fig.\ref{Fig1}(c) shows the same setup like in Fig.\ref{Fig1}(b) but now for an off-diagonal
coupling that is stronger within an LS-domain as compared to between LS-domains.
As a result the localization onto the LS-domains becomes even more pronounced: essentially all
eigenstates show this tendency with different degrees of expression. 

\noindent
In spite of the above-exemplified localization properties of eigenstates in the presence of LS-domains
and their importance for various setups \cite{Roentgen19,Morfonios20} a more profound
understanding beyond the degenerate perturbation theoretical approach developed in
\cite{Roentgen19} is still lacking. In the present work we address this gap and will develop relevant insights 
concerning the mechanism for localization in the presence of local symmetries thereby focusing
on local reflection symmetries. In section \ref{sec:wce}
we perform a weak coupling expansion for the eigenvectors based on a closed form expression provided in
ref.\cite{Parlett98}. The resulting eigenvalues and eigenvectors provide us with first indications of the localization
properties for small coupling values. In section \ref{sec:dsl} we extend this argument on basis of the 
fact that subdomains related by a local symmetry operation are isospectral and provide us with pairwise degenerate states that
are coupled via the intersubdomain coupling. We provide an evidence-based analysis of the eigenvalue splitting 
with varying inter-subdomain couplings and discuss the case of delocalization within a (sub-)domain versus the localization on
LS-domains. Our conclusions and outlook are provided in section \ref{sec:concl}.

\section{Weak coupling expansion} \label{sec:wce}

\noindent
We will focus in the following on a TB Hamiltonian which takes on the following appearance

\begin{equation}
{\cal{H}} = \sum_{i=1}^{N} a_i |i \rangle \langle i| + \sum_{ \langle i,j \rangle} \epsilon_{ij} |i \rangle \langle j| 
\label{eq1}
\end{equation}

\noindent
We assume, if not explicitly state otherwise, that the off-diagonal coupling between nearest neighbors 
$\langle i,j \rangle$ is constant i.e. $\epsilon_{ij}=\epsilon$ for the complete chain of length $N$. 
Our choice of the corresponding onsite energies $a_i$ will reflect the absence or presence of local symmetries.
Explicit analytical expressions for the eigenvalues and eigenvectors of TB Hamiltonians
\cite{Kouachi06,Kulkarni99,Willms08,Noschese13,Nakatsukasa12} are known only for special cases,
such as tridiagonal Toeplitz matrices, and do not cover the above-given general Hamiltonian (\ref{eq1}).
However, in ref.\cite{Parlett98} expressions for the squared components 
$s_{\mu i}^2$ of the normalized eigenvectors have been provided which read as follows

\begin{equation}
s_{\mu i}^2 = \chi_{1:\mu-1}(\lambda^{i}) \chi_{\mu+1:N}(\lambda^{i}) / \chi^{\prime}_{1:N} (\lambda^{i})
\label{eq2}
\end{equation}

\noindent
where $\mu$ and $i$ refer to the eigenvector component and the label of the eigenstate/eigenenergy.
$\chi$ represents the characteristic polynomial (CP). $\chi^{\prime}_{1:N}$ is the derivative of the CP of the complete
Hamiltonian matrix. $\chi_{1:\mu-1}(\lambda^{i})$ is the CP of the $(\mu -1) \times (\mu -1)$ submatrix
obtained by deleting the rows and columns labeled by $\mu,....,N$ of the complete matrix
and taken at the eigenvalue $\lambda^{i}$. Correspondingly for $\chi_{\mu+1:N}(\lambda^{i})$.
While eq.(\ref{eq2}) represents a remarkable and distinguished result it is of implicite character,
in the sense that both the knowledge of the exact eigenvalues $\lambda^{i}$ and the characteristic
polynomial $\chi$ are needed to evaluate this expression. To nevertheless exploit the relationship
(\ref{eq2}) we will perform a weak coupling expansion of both the eigenvalues $\lambda^{i}$ and the
CPs $\chi$. This will render the above expression (\ref{eq2}) explicite and of
direct use. Since we focus on a second order expansion w.r.t. the coupling strength $\epsilon$
the validity of our approach (within this section) is restricted to small values of $\epsilon$,
i.e. weak couplings. This way we will be able to see the onset of the localization behaviour due to local
symmetries whereas an extension of it will be addressed in the following section.

\noindent
We expand up to second order in the coupling $\epsilon$ (around $\epsilon = 0$) which yields for the eigenvalues
$\lambda^{i} \approx \lambda^{i}_0 + \epsilon \lambda^{i}_1 + \epsilon^2 \lambda^{i}_2$
and $\lambda^{i}_0 = a_i$. For the following derivations we use extensively the recursion relation
for the CP \cite{Parlett98} providing the CP of the $j-$dimensional matrix as a function of the CPs of the 
$(j-1)-$ and $(j-2)-$dimensional matrices which reads

\begin{equation}
\chi_{1:j}(\lambda) = \left( \lambda - a_j \right) \chi_{1:j-1}(\lambda) - \epsilon^2 \chi_{1:j-2} 
\label{eq2a}
\end{equation}

\noindent
After some algebra, one obtains an explicite expression for the CP valid up to second order of $\epsilon$

\begin{widetext}
\begin{eqnarray} \label{eq3}
\chi_{1:\mu-1} (\lambda^{i}) = \prod_{j=1}^{\mu-1} X_j^{i} + \epsilon \lambda_1^{i}
\left( \sum_{j=1}^{\mu-1} \prod_{\substack{k=1\\ k \ne j}}^{\mu-1} X_k^{i} \right) 
+ \epsilon^2 \left( \lambda_2^{i} \sum_{j=1}^{\mu-1} \prod_{\substack{k=1\\ k \ne j}}^{\mu-1} X_k^{i} 
+ {\lambda_1^{i}}^2 \sum_{k=1}^{\mu-2} \sum_{\substack{l=1\\ l>k}}^{\mu-1} \prod_{\substack{r=1\\ r \ne k,l}}^{\mu-1} X_r^{i}
 - \sum_{j=1}^{\mu-2} \prod_{\substack{l=1\\ l \ne j,j+1}}^{\mu-1} X_l^{i} \right)
\end{eqnarray}
\end{widetext}

\noindent
where $X_l^{i} = (\lambda^{i}_0-a_l)$ with $\lambda^{i}_0 = a_i$.
A corresponding expression holds for $\chi_{\mu+1:N}(\lambda^{i})$ in eq.(\ref{eq2}).
Note that in the above-involved derivation additional boundary terms appear for the components $\mu \le 3, \mu \ge N-2$
which we safely ignore for the following line of argumentation. The denominator $\chi^{\prime}_{1:N} (\lambda^{i})$
in eq.(\ref{eq2}) can be expressed as

\begin{equation}
\chi^{\prime}_{1:N} (\lambda^{i}) = \sum_{\mu=1}^{N} 
\chi_{1:\mu-1}(\lambda^{i}) \chi_{\mu+1:N}(\lambda^{i})
\label{eq4}
\end{equation}

\noindent
where applies $\chi_{1:0}=\chi_{N+1:N}=1$. Inserting eqs.(\ref{eq3},\ref{eq4}) and the corresponding expression
for $\chi_{\mu+1:N}(\lambda^{i})$ in eq.(\ref{eq2}) and providing the expansion of the inverse up to second
order of $\epsilon$ yields for the square of the eigenvector components 

\begin{widetext}
\begin{eqnarray} \label{eq5}
s_{\mu i}^2 &=& \left( \frac{{\cal{D}}_{\mu}^{(i,0)}}{\sum_{\nu=1}^{N} {\cal{D}}_{\nu}^{(i,0)}} \right)
+ \epsilon \left( \sum_{\nu=1}^{N} {\cal{D}}_{\nu}^{(i,0)} \right)^{-2} 
\left( {\cal{D}}_{\mu}^{(i,1)} \sum_{\nu=1}^{N} {\cal{D}}_{\nu}^{(i,0)}
- {\cal{D}}_{\mu}^{(i,0)} \sum_{\nu=1}^{N} {\cal{D}}_{\nu}^{(i,1)} \right) \\ \nonumber
&+& \epsilon^2 \left( \sum_{\nu=1}^{N} {\cal{D}}_{\nu}^{(i,0)} \right)^{-2} 
\left( {\cal{D}}_{\mu}^{(i,2)} \sum_{\nu=1}^{N} {\cal{D}}_{\nu}^{(i,0)}
- {\cal{D}}_{\mu}^{(i,1)} \sum_{\nu=1}^{N} {\cal{D}}_{\nu}^{(i,1)} \right) \\ \nonumber
&+& \epsilon^2 \left( {\cal{D}}_{\mu}^{(i,0)} \right)
 \left( \sum_{\nu=1}^{N} {\cal{D}}_{\nu}^{(i,0)} \right)^{-3} 
\left( \left(\sum_{\nu=1}^{N} {\cal{D}}_{\nu}^{(i,1)}\right)^2
- \left( \sum_{\nu=1}^{N} {\cal{D}}_{\nu}^{(i,0)} \right)
\left(\sum_{\nu=1}^{N} {\cal{D}}_{\nu}^{(i,2)} \right) \right)
\end{eqnarray}
\end{widetext}

\noindent
where

\begin{widetext}
\begin{eqnarray} \label{eq6}
{\cal{D}}_{\mu}^{(i,0)} &=& \prod_{\substack{j=1\\ j \ne \mu}}^{N} X_j^{i} \hspace*{1cm}
{\cal{D}}_{\mu}^{(i,1)} = \lambda_1^{i} \left( \sum_{\substack{j=1\\ j \ne \mu}}^{N} 
\prod_{\substack{k=1\\ k \ne j,\mu}}^{N} X_k^{i} \right) \\ \nonumber
{\cal{D}}_{\mu}^{(i,2)} &=& {\lambda_1^{i}}^2 \left(
\left(\sum_{j=1}^{\mu-1} \prod_{\substack{k=1\\ k \ne j}}^{\mu-1} X_k^{i} \right)
\left(\sum_{j=\mu+1}^{N} \prod_{\substack{k=\mu+1\\ k \ne j}}^{N} X_k^{i} \right)
+ \left(\sum_{k=1}^{\mu-2} \sum_{\substack{l=1\\ l > k}}^{\mu-1}
+ \sum_{k=\mu+1}^{N-1} \sum_{\substack{l=\mu+1\\ l > k}}^{N} \right)
\prod_{\substack{j=1\\j\ne \mu,l,k}}^{N} X^{i}_{j} \right) \\ \nonumber
& & + \lambda_2^{i} \left(\sum_{\substack{j=1\\ j\ne \mu}}^{N} \prod_{\substack{k=1\\ k \ne j,\mu}}^{N} X_k^{i}
\right) - \left( \sum_{\substack{j=1\\ j\ne \mu-1,\mu}}^{N-1} \prod_{\substack{k=1\\ k \ne \mu,j,j+1}}^{N} X_k^{i} \right)
\\ \nonumber
\end{eqnarray}
\end{widetext}

\noindent
The first and second order contributions $\lambda^{i}_k; k = 1,2;i=1,...,N$ to the eigenvalues $\lambda^i$ are 
derived on basis of eqs.(\ref{eq2a},\ref{eq3}). The CP has been expanded up to second order of $\epsilon$
by using the above recursion relation iteratively and consequently we
equate it to zero. We then demand that the terms of each order of $\epsilon$, namely the 'coefficients' of
$\epsilon$ and $\epsilon^2$ are zero resulting on equations for $\lambda^{i}_k; k = 1,2;i=1,...,N$.

\noindent
Let us first focus on the 
case of a nondegenerate zeroth order spectrum, i.e. all the diagonal elements are pairwise different
$a_i \ne a_j~\text{for}~i \ne j$, which corresponds to a situation whose eigenstate map is shown in Fig.\ref{Fig1}(a).
We obtain $\lambda^i_{1} = 0~\forall~i$ and only the second order contribution is nonzero. It reads

\begin{equation}
\lambda^i_{2} = \left(X_{i+1}^{i}\right)^{-1} + \left(X_{i-1}^{i}\right)^{-1}
\label{eq7}
\end{equation}

\noindent
For the corresponding eigenvector one has to distinguish between the original zeroth order site $i$ and the
neighboring ones. We therefore obtain 

\begin{eqnarray}
s_{ii}^2 &=& 1 - \epsilon^2 \left({\cal{D}}_i^{(i,0)} \right)^{-1} \left( \sum_{\substack{\nu=1, \nu \ne i}}^{N}
{\cal{D}}_{\nu}^{(i,2)} \right) \\ \nonumber
s_{\mu i}^2 &=& \epsilon^2 \left({\cal{D}}_{i}^{(i,0)}\right)^{-1} \left({\cal{D}}_{\mu}^{(i,2)}\right), \hspace*{0.2cm}
 \mu \ne i
\label{eq8}
\end{eqnarray}

\noindent
This demonstrates, as expected, that primarily the zeroth order site is dominantly contributing and the
neighboring sites appear in second order of the coupling $\epsilon$ as shown in
Fig.\ref{Fig1}(a). The exact eigenstates possess of course
contributions of higher order w.r.t. $\epsilon$ and are slightly more delocalized. 

\noindent
Let us now address the case of two neighboring sites $i,i+1$ that possess the same on-site energies
for zero coupling and we focus on
the eigenvalues of the TB-Hamiltonian which emerge from this degenerate pair when increasing the coupling strength.
Such a configuration appears if an LS-domain based
on local reflection is present in the Hamiltonian, as discussed above. We therefore have
$\lambda^{i}_{0}= \lambda^{i+1}_{0} = a_i = a_{i+1}$ relating to the two central sites
of the LS-domain. From the CP we obtain then 
$\lambda_{1}^{i} = -\lambda_{1}^{i+1} = 1$ i.e. naturally a linear order term appears in the expansion of the
eigenvalue(s). Due to the fact that ${\cal{D}}_{\mu}^{(i,0)} = 0~\forall~\mu$ the expansion in the nominator 
and the denominator of eq.(\ref{eq2}) start in linear order of $\epsilon$, respectively, and 
we obtain for the eigenvectors

\begin{widetext}
\begin{subequations}
\begin{eqnarray} 
s_{\mu \alpha}^2 &=& \frac{1}{2} + \epsilon \left( \sum_{\nu=1}^{N} {\cal{D}}_{\nu}^{(\alpha,1)} \right)^{-2} 
  \left( {\cal{D}}_{\mu}^{(\alpha,2)} \left( \sum_{\nu=1}^{N} {\cal{D}}_{\nu}^{(\alpha,1)} \right)
- {\cal{D}}_{\mu}^{(\alpha,1)} \left( \sum_{\nu=1}^{N} {\cal{D}}_{\nu}^{(\alpha,2)} \right) \right)
\hspace*{0.2cm} \text{for} \hspace*{0.1cm} \alpha,\mu \in \left[i,i+1\right] 
\label{eq9a} \\
s_{\mu \alpha}^2 &=& \epsilon \left( \sum_{\nu=1}^{N} {\cal{D}}_{\nu}^{(\alpha,1)} \right)^{-1}
 {\cal{D}}_{\mu}^{(\alpha,2)}  \hspace*{0.2cm} \text{for} \hspace*{0.1cm} 
\alpha \in \left[i,i+1\right], \mu \notin \left[i,i+1\right]
\label{eq9b}
\end{eqnarray}
\end{subequations}
\end{widetext}

\noindent
Eqs.(\ref{eq9a},\ref{eq9b}) shows that the central two sites with equal on-site energy possess in leading order
the same values $\frac{1}{2}$ for the corresponding squared eigenvector components. This can be interpreted
as follows. The degenerate zeroth order pair $i,i+1$ with on-site values $a_i=a_{i+1}$ forms a 'nucleus'
for the localization of the corresponding eigenvectors on the LS-domain. Moving outward from this nucleus within the LS-domain 
further next order contributions to the corresponding eigenvector components appear, see eq.(\ref{eq9b})
that have a corresponding counterpart for the central components, see eq.(\ref{eq9a}) \cite{footnote1}. As far as we can judge
within the weak coupling expansion up to second order $\epsilon^2$ the central on-site energy degeneracy
represents the seed for the localization of the eigenvectors on the LS-domains. In the next section we will
explore the more general case of higher-order effects and/or stronger couplings based on the argument
of a pairwise degeneracy of all eigenstates within an isolated LS domain. This will
provide us with even stronger arguments for the localization mechanism at work.
A final note is in order concerning the above weak coupling expansion. In principle, a higher order
expansion follows the same procedure. However, the corresponding algebraic manipulations become tedious and,
therefore, a computer algebraic approach might be more feasible.

\begin{figure*}[t]
\centering
\includegraphics[width=\linewidth]{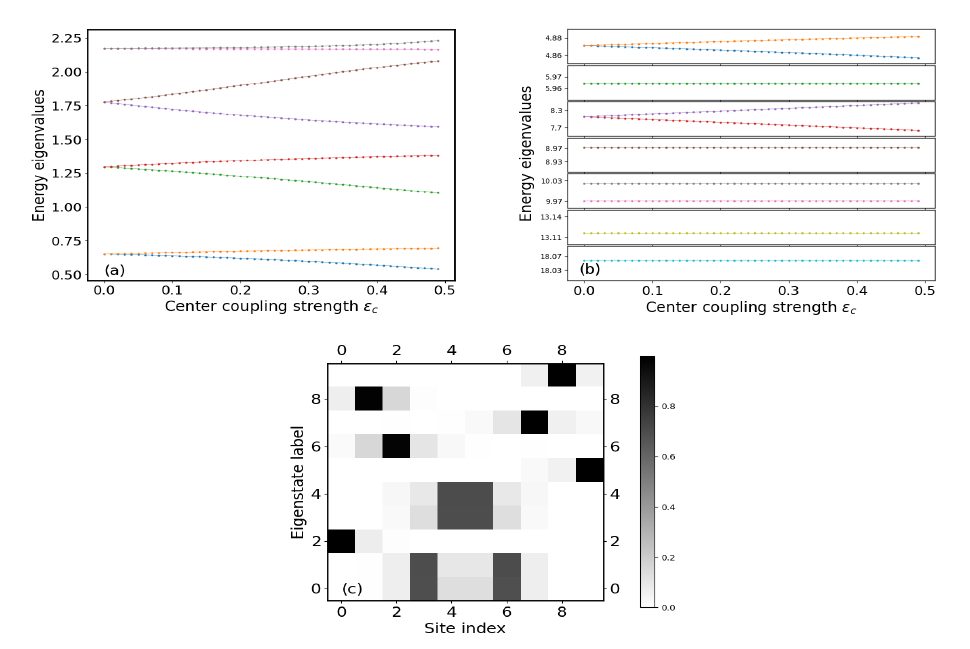}
\caption{Energy eigenvalue spectra (a,b) for a varying center coupling value $\epsilon_c = 0-0.5$. 
(a) A single reflection LS-domain consisting of 8 sites and a subdomain coupling strength 
$\epsilon = 0.4$. Diagonal values of the TB-Hamiltonian are $1.9,1.4,1.1,1.5,1.5,1.1,1.4,1.9$.
We observe the pairwise degeneracy of the eigenvalues for $\epsilon_c=0$ and their linear
splitting with increasing value of $\epsilon_c$ (note that this splitting is very small for the
pair with the largest eigenvalues and barely visible on the scale of the figure).
(b) A stack of subfigures showing the eigenvalue spectrum for a single reflection LS-domain embedded
into an asymmetric environment. Diagonal values of the
10-dimensional TB-Hamiltonian are $6.0,13.0,10.0,5.0,8.0,8.0,5.0,10.0,18.0,9.0$ containing a 6-dimensional
reflection symmetric domain. As for (a) the center coupling obeys $\epsilon_c = 0-0.5$ and the remaining
couplings are $\epsilon = 0.5$. (c) The eigenstate map of the setup (b) for $\epsilon_c = \epsilon = 0.5$.
Eigenvector components are shown with varying site index on the horizontal axis, and sorted w.r.t. increasing energy eigenvalues
from bottom to top along the vertical axis.
\label{Fig2}}
\end{figure*}

\section{Degenerate subspace localization} \label{sec:dsl}

\noindent
Let us now consider a broader range of values for the coupling $\epsilon$ but still staying
in the regime of a strong contrast $\frac{a_i-a_{i+1}}{\epsilon}>1 \hspace*{0.1cm} \text{for} \hspace*{0.1cm} a_i \ne a_{i+1}$. 
If the contrast would be significantly smaller than one the tendency of delocalization
for all eigenstates would smear them out and, consequently, the impact of the LS-domains on the localization
becomes negligible. Naturally, since analytical calculations are not possible in this
regime, we base our analysis on the results of numerical simulations.
To begin with we focus  on a single LS-domain consisting of two subdomains that are mapped onto
each other by a (local) symmetry operation and we assume (as always) open boundary conditions.
Let us assume that the center coupling (i.e. intersubdomain coupling) between the two subdomains is zero, i.e. $\epsilon_c = 0$.
In case of a translation the two submatrices belonging
to the two subdomains are identical and in particular isospectral, i.e. they possess identical
eigenvalues. Of course, they also possess the same eigenvectors. This isospectrality holds also for a reflection operation.
The latter can be proven by performing $\lfloor \frac{N}{2} \rfloor$ exchange operations of the
columns and subsequently of the rows from the outside to the inside of the original submatrix 
which yields the submatrix of the symmetry transformed subdomain and makes up
for a total sign change of $(-1)^{2 \cdot \lfloor \frac{N}{2} \rfloor}=+1$ of the corresponding
determinant. The corresponding eigenvector components for the two subdomains 
are related by $s_{ji} = s_{N-j+1 \hspace*{0.05cm}i}$.
The above implies that the spectrum consists of pairs of degenerate eigenvalues for $\epsilon_c = 0$, independent
of the strength of the intra-subdomain coupling $\epsilon$. With increasing value of the
coupling $\epsilon$ the subdomain eigenstates become increasingly delocalized inside the subdomain.

\noindent
The center coupling $\epsilon_c$ represent a key quantity for the analysis of the spectral and
eigenstate properties since it couples isospectral submatrices and, importantly, since it
maintains the symmetry properties of the setup while being varied.

\begin{figure*}[t]
\centering
\includegraphics[width=\linewidth]{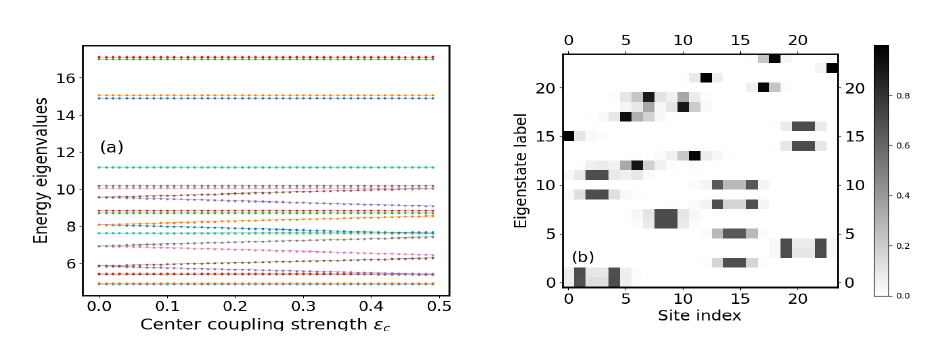}
\caption{A 24-dimensional TB-Hamiltonian consisting of four different 6-dimensional neighboring reflection-based LS-domains.
The four domains involve the sites $[0-5],[6-11],[12-17],[18-23]$.
(a) Energy eigenvalue spectra with varying (all four) center coupling strengths $\epsilon_c = 0-0.5$.
(b) The corresponding eigenstate map for $\epsilon = \epsilon_c = 0.5$.
\label{Fig3}}
\end{figure*}

\noindent
As a consequence of the above discussed properties, we observe that all pairwise degenerate eigenvalues of
the decoupled subdomains split linearly
if we increase their center coupling strength $\epsilon_c$ starting from a zero value. Fig.\ref{Fig2}(a) shows
an 8-dimensional case for the on-site values $1.9,1.4,1.1,1.5,1.5,1.1,1.4,1.9$ and $\epsilon = 0.4$.
For a given on-site energy configuration the strength of the splitting increases with an increasing value 
for the subdomain coupling strength $\epsilon$. For larger values of $\epsilon_c$ a nonlinear behaviour
takes over. Let us now bring a reflection symmetric domain in contact with an asymmetric environment.
The resulting eigenvalue spectrum is shown in Fig.\ref{Fig2}(b) for the case of a 
6-dimensional reflection symmetric domain with diagonal values $10.0,5.0,8.0,8.0,5.0,10.0$ 
in contact with 2 left and 2 right attached asymmetric sites. There is two main effects we consequently observe.

\noindent
First the symmetry breaking leads to an energy splitting of the originally 
degenerate pair for $\epsilon_c = 0$ which is tiny for the central neighboring sites
but increases significantly for pairs located closer to the boundaries of the LS domain.
Second there is still large intervals of $\epsilon_c$ for which the two inner of the three pairs 
of the LS-domain show an approximate linear splitting and dependence on $\epsilon_c$.
The other eigenvalues show a behaviour which is (on the shown scale) insensitive to the variations of $\epsilon_c$.
Fig.\ref{Fig2}(c) shows for the same setup
like in Fig.\ref{Fig2}(b) and for $\epsilon_c = \epsilon = 0.5$ the corresponding eigenstate map.
There is four low energy eigenstates which are localized on the LS-domain and show dominant
values on two sites. These correspond to the eigenvalues that show a linear splitting in Fig.\ref{Fig2}(b).
The other eigenstates show a dominant single site behaviour, as ubiquitous in the absence
of any (local) symmetry. 

\noindent
Note that the coupling $\epsilon$ in the subdomains plays an important
role in the above discussed behaviour: it helps delocalizing the eigenstates on the subdomain
leading to a stronger sensitivity w.r.t. the center coupling $\epsilon_c$ and consequently an enforced
localization on the LS-domain overall. This argument has to be taken with a grain of salt, since making
the coupling too strong yields an overall delocalization not 'seeing the particular imprint' of the LS-domain at all.

\noindent
Fig.\ref{Fig3}(a,b) show the eigenvalue spectrum and an eigenstate map for a more complex setup: a 
24-site Hamiltonian with 4 different 6-dimensional reflection-symmetric domains. 
We observe again a substantial subset of linearly splitting eigenvalues (for some of them the
splitting is not visible on the scale of Fig.\ref{Fig3}(a)) with varying all center couplings $\epsilon_c$. 
The corresponding eigenstate map in Fig.\ref{Fig3}(b) shows that 14 out of the 24 eigenstates are
confined to LS-domains and follow their characteristic profile.

\section{Conclusions and outlook} \label{sec:concl}

\noindent
Breaking global symmetries and retaining them locally leads directly to a plethora of possible
setups that fall into the gap between perfect order and complete disorder. The quest for the
characterization of the properties of such systems is of immediate interest in view of the fact
that there are many different ways of implementing local symmetries: they could be e.g. isolated,
neighboring, overlapping and covering or non-covering in terms of their domains. 
In the present work we make an attempt to better understand the eigenstate localization behaviour of
locally symmetric Hamiltonians which has been observed in previous works such as refs.\cite{Roentgen19,Morfonios17}.
In those works it has consistently been monitored that a certain significant portion of the eigenstates prefers
to localize, i.e. have their dominant eigenvector components, on locally symmetric domains of the
total setup. Therefore one can use the freedom of incorporating local symmetries in a 'device'
to steer the localization behaviour of the corresponding eigenstates.

\noindent
We have been focusing on local reflection symmetries to analyze the spectrum of eigenvalues and
eigenstates. Our first analysis step was based on the closed form expressions for the eigenvectors
of tight-binding Hamiltonian provided in ref.\cite{Parlett98}. While these expressions involve
the exact characteristic polynomial and eigenvalues a weak coupling expansion renders them of immediate
use within a low order series approximation of the coupling strength. If no local
symmetries are present we obtain the expected single site dominant behaviour for the eigenvectors,
with some small amplitude on neighboring sites. It is the degeneracy of the zeroth order eigenvalues
at the center of a locally reflection symmetric domain which serves as a nucleus for the eigenvector delocalization
on locally symmetric domains in the weak coupling limit. 

\noindent
To address the regime of larger values for the couplings, but still remaining within the range
of a strong contrast, we numerically explored for several examples the spectral behaviour of the eigenvalues
and their eigenstate localization in the presence of one or several local symmetries.
Our first observation was the fact that mapping a subdomain via a reflection operation onto its image
subdomain is an operation which maintains the underlying eigenvalues of the corresponding Hamiltonian,
i.e. the two subdomain Hamiltonian are isospectral independently of the value of the coupling strength
within the subdomain. As a consequence, for zero center coupling between the subdomains
we encounter degenerate pairs of eigenvalues. Switching on the center coupling leads to a linear splitting
of all pairwise degenerate eigenvalues if there is no coupling to an (asymmetric) environment, i.e. to neighboring sites.
The case of interest is the one for which a locally symmetric domain is embedded into a larger setup.
Then it is only a subset of the eigenvalues which split linearly, depending on the strength of the
coupling to the environment. This holds for each locally symmetric domain separately. Our analysis of the
eigenstates via eigenstate maps reflects the correlation between the linear splitting and the
localization of the eigenstates on domains with local symmetries. The coupling to the environment 
weakens the localization tendency on the underlying locally symmetric domains: from the surface of the domains
to their center this process becomes less pronounced.

\noindent
While the above facts illuminate the origin and mechanism of the ubiquitously-observed localization
of eigenstates in the framework of setups exhibiting local symmetries, it also opens up new perspectives.
In our approach the key ingredient is the isospectrality of the two subdomains constituting the locally symmetric
domain. One could now choose as subdomains not symmetry related ones but an original subdomain and a similarity
transformed one. By construction, both subdomains possess the same set of eigenvalues. It is then an open question
to what extent and under what (additional) criteria the coupled subdomains would show, once embedded into an
environment, corresponding localization properties. In this context it is an intriguing perspective to explore the recently
investigated latent or hidden symmetries \cite{Roentgen21,Morfonios21,Roentgen23} which are based on
isospectral reduction techniques of spectral graph theory \cite{Smith19,Bunimovich14,Kempton20} thereby
generalizing the here studied local symmetries.

\section{Acknowledgments} \label{sec:acknowledgments}

P.S. acknowledges fruitful discussions with J. Schirmer on degenerate perturbation
theory and with M. R\"ontgen on the impact of local symmetries.
This work has been supported by the Cluster of Excellence “Advanced Imaging of Matter” of the Deutsche
Forschungsgemeinschaft (DFG)-EXC 2056, Project ID No. 390715994.


\begin{thebibliography}{99}
\bibitem{footnote1} We remark that our line of argumentation is valid independently of the definite order 
$\mathcal{O}({\epsilon^n})$ which appears to be nonzero besides the constant value $\frac{1}{2}$ in eqs.(\ref{eq9a},
\ref{eq9b}).
\end{thebibliography}
\end{document}